\documentclass[conference]{IEEEtran}

\usepackage[bordercolor=white]{todonotes}
\usepackage{subfigure}
\usepackage{paralist}
\usepackage{balance}
\usepackage[hyphens]{url}
\usepackage[utf8]{inputenc}

\newcommand\secref{Section~\ref}

\begin{document}

\title{The Society Spectrum:\\ Self-Regulation of Cellular Network Markets}

\author{
    \IEEEauthorblockN{Patrick
      Zwickl\IEEEauthorrefmark{1}, Ivan Gojmerac\IEEEauthorrefmark{2}, Paul Fuxjaeger\IEEEauthorrefmark{3}, Peter Reichl\IEEEauthorrefmark{1}, Oliver Holland\IEEEauthorrefmark{4} } %
    \IEEEauthorblockA{
      \IEEEauthorrefmark{1} Faculty of Computer Science, University of Vienna, Austria, E-mail: \{patrick.zwickl$\vert$peter.reichl\}@univie.ac.at  }
      \IEEEauthorblockA{
      \IEEEauthorrefmark{2}Digital Safety and Security Department, AIT -- Austrian Institute of Technology, Vienna, Austria \\ E-mail: ivan.gojmerac@ait.ac.at} 
    \IEEEauthorblockA{
    \IEEEauthorrefmark{3}FTW -- Telecommunications Research Center Vienna, Vienna, Austria, E-mail: fuxjaeger@ftw.at }
     \IEEEauthorblockA{
    \IEEEauthorrefmark{4} Centre for Telecommunications Research, King’s College London, UK, E-mail: oliver.holland@kcl.ac.uk }
  }


\maketitle

\begin{abstract}


Today’s cellular telecommunications markets require continuous monitoring and intervention by regulators in order to balance the interests of various stakeholders.  In order to reduce the extent of regulatory involvements in the day-to-day business of cellular operators, the present paper proposes a ``self-regulating'' spectrum market regime named ``society spectrum''. This regime provides a market-inherent and automatic self-balancing of stakeholder powers, which at the same time provides a series of coordination and fairness assurance functions that clearly distinguish it from ``spectrum as a commons'' solutions. The present paper will introduce the fundamental regulatory design and will elaborate on mechanisms to assure fairness among stakeholders and individuals. This work further puts the society spectrum into the context of contemporary radio access technologies and cognitive radio approaches.




\end{abstract}


\IEEEpeerreviewmaketitle

\section{Introduction}
\label{introduction}

In economics, distributed equilibria and especially self-regulating systems are preferable over centralized mechanisms due to their simplicity. Today's telecommunications market is, however, a good way off this ideal situation. The regulator needs to actively monitor and moderate markets in order to not only balance the market powers of consumers and providers but also among smaller and larger providers. With the emergence of and controversies around Net Neutrality, the regulator's role has substantially been extended towards monitoring discriminatory actions on the content axis---either due to \textit{foreclosure} (excluding measures) or \textit{sabotage} (quality degradations)---following Ofcom's taxonomy in \cite{croicioni}). This view has been supported by recent side payment deals from data hungry content providers such as Netflix\footnote{Netflix is currently issuing side payments to Comcast\footnote{\url{http://time.com/80192/netflix-verizon-paid-peering-agreement/}, last accessed: \today} in order to provide the expected service quality.} to incumbent network operators such as Comcast or Verizon.

Along the lines of other industries' experiences, e.g., aviation mergers \cite{borenstein1990airline}, we observe a direct relationship between the number of directly competing network operators and the attractivity of their product offerings\footnote{Prices have sharply increased following a recent merger of cellular network operators in Austria. \url{http://www.ft.com/cms/s/0/dba557c8-9c91-11e3-b535-00144feab7de.html#axzz3RQPEsdk3}. last accessed: \today} (price, quality, etc.). Hence, operators tend to seek for market consolidation, while consumers desire a more fragmented market. 
	

	
	Another battle ground \cite{zwickl2013wi} emerges around spectrum competition constellations in the unlicensed spectrum and its intermingling with licensed technologies (i.e., LTE-U) that aim at competing with Wi-Fi.
	 
	 Under these circumstances, the work of regulatory bodies has become tremendously difficult, aiming to support (at the same time):
	 \begin{enumerate}
	    \item Incentives to invest in state-of-the-art infrastructure covering the majority of the population and the considered geographical area;
	    \item The creation of reasonable market pressures protecting the interests of consumers, which includes mechanisms to avoid typical market failures and markets issues such as collusion, too high or too low competition, myopic investment decisions, etc.;
	    \item Fairness among users but also competitors;
	    \item The assurance of basic Internet connectivity in order to lower the digital divide; 
	    \item The availability assurance of ubiquitous high quality Internet, e.g., meeting the demands of industrial use cases (e.g., with respect to \textit{Industry 4.0}); and
	    \item The avoidance of overregulation that hampers market mechanisms to properly function.
	 \end{enumerate}

When concentrating on radio access and spectrum markets, neither today's unlicensed (e.g., Wi-Fi) nor licensed spectrum (e.g., LTE) solutions entirely meet those standards without active regulatory intervention all the way from the operational to the strategic level. Unlicensed spectrum still lacks in coordination and incentives to massively deploy infrastructure also in rural areas, which may lead to coverage, quality and fairness issues, and also may entail colliding commercial interests (see \cite{zwickl2013wi}). The licensed spectrum on the other hand is strictly regulated and requires constant intervention to assure its functioning. Wrong decisions (such as during spectrum auctions) may not be able to be corrected for the subsequent years, which appears overly complex and inflexible. 

For these reasons, the present paper proposes the establishment of a novel agency-assisted self-regulated \cite{ogus1995rethinking} and highly transparent spectrum market that meets all defined goals. This market approach takes the positive characteristics of both the unlicensed and licensed spectrum approaches in order to minimize the required regulatory interventions, while still almost perfectly balancing all stakeholders' interests in the long run. This goal is targeted by proposing the concept of a  \textit{society spectrum} as a relief spectrum for market pressure fluctuations that may harm the sustainability of the entire ecosystem from the consumers' or the operators' point of view. The society spectrum will automatically counteract insufficiently functioning markets arising due to collusion, insufficient competition, too high competition, bargaining powers (e.g., due to managing the access to scarce spectrum), etc.

	 	The remainder of this work is structured as follows: After reviewing related works in \secref{sec:related}, a structural separation of markets powers is presented in \secref{sec:fundamentals} as prerequsite for a clean market design. The core contribution, i.e., a self-regulating market regime, is presented in \secref{sec:market_design}. Concepts assuring fairness among users and operators are introduced in \secref{sec:fairness} and associated to a radio and access network technology perspective. The access to a fair share of the spectrum is translated to the operator landscape in \secref{sec:spectrum_access}, after which the paper is closes with concluding remarks in \secref{conclusions}.
	 	
	 	

\section{Related Work}
\label{sec:related}

According to \cite{kpmg96} and \cite{blackman1998convergence}, regulators play a dual role as preventers of market failures and as bodies acting in public interest\footnote{While in \cite{blackman1998convergence} the argumentation is mainly part of a content-centric analysis, the abstract separation of concepts is still absolutely valid for the context of spectrum regulation.}. Regulation may ``occur in the three traditional components of separation of powers: legislation, enforcement, and adjudication''  where particular industries may be regulated in any of those components \cite{swire1997markets}. Self-regulation as opposed to ``command and control'' \cite{sinclair1997self} (i.e., classical regulation) has been a widely discussed topic spanning multiple industries and research directions, e.g., see \cite{sinclair1997self,ogus1995rethinking,king2000industry,swire1997markets}. The definitions for self-regulation are widely diverging---e.g., Swire \cite{swire1997markets}  interprets self-regulation as set of self-imposed industry guidelines and comparable measures (subject to ``industry morality'' \cite{king2000industry}), while Ogus \cite{ogus1995rethinking} defines a series of self-regulation flavors ranging from free market approaches to strict policies. For clarity, the present paper will use self-regulation in the sense of limiting the required regulatory interventions (``command and control'') to a minimum, especially directing regulatory measures more towards longer term strategic decisions instead of close involvements in day-to-day operations. However, similarly to \cite{blackman1998convergence}, we acknowledge that an ``Unconstrained Self-regulation'' (as defined in \cite{ogus1995rethinking}) would be irresponsible when acting upon public interest. 

In 1980, Melody \cite{melody1980radio} in particular described spectrum as ``unique natural resource’’ that contrary to other natural resources would require cooperation. Melody further stressed that the spectrum allocation will likely involve administrative processes in assistance to market mechanisms. Faulhaber \& Farber \cite{faulhaber2003spectrum} added that spectrum as \textit{commons} does not work under scarcity, that is why ``ownership models’’ are required (denoted by exclusive spectrum in this work). Faulhaber \& Farber have further foreseen the dynamic and automated realtime selling of exclusively owned spectrum and favor the intermingling of spectrum as ``commons'' with exclusive spectrum, without detailing a new market setup. Recent research on Cognitive Radio (CR) has also focused on developing means for the more dynamic spectrum usage and economic utilization \cite{niyato2008market}. This idea is often referred to as Dynamic Spectrum Access (DSA), which has received substantial attention in technical, regulatory, and economic research so far, e.g., in \cite{chapin2007,rodriguez2005market}. Chapin \textit{et al.} for example define triggers (``available spectrum’’, ``customer demand’’, ``low transaction costs’’) for creating ``liquidity'' in the dynamic access of spectrum and market opportunities for DSA-specific services. While regulatory certainty about the issuing and usage of secondary spectrum is required, the risk of interferences has to be minimal \cite{chapin2007}. Optimal spectrum pricing for CDMA is further discussed in \cite{rodriguez2005market}.  Applicable to CR/DSA scenarios, Holland et al. define the Pluralistic Licensing concept to improve spectrum flexibility and autonomy in licensing \cite{holland2012pluralistic}. Forde \textit{et al.} \cite{forde2011exclusive} have further envisioned a more dynamic spectrum assignment enabled both via a virtualization of Radio Access Networks (RANs) and especially due to the technological advancements around CRs, Software-Defined Networking (SDN) and Long Term Evolution (LTE) Advanced. In their approach, they transfer essential RAN functions to the cloud in order to better integrate Mobile Virtual Network Operators (MVNOs).



Thus we conclude that spectrum represents a unique value for the exclusive holders and that even more restricted secondary usage scenarios are of interest. Moreover, self-imposed industry guidelines will be insufficient to target the complexity of the telecommunications market, especially when assuming limited industry morality (i.e., a useful common assumption). The present work hence aims at establishing a system-inherent self-regulation based on a redesigned spectrum market, which is enacted by regulatory bodies. Following the position in \cite{faulhaber2003spectrum} the present work intermingles spectrum markets as commons and exclusively operated alternatives in order to dynamically react to market conditions, and hence to reduce the need for regulatory interventions. 


\section{Separation of Powers} \label{sec:fundamentals}

The regulatory primary interest is to assure a sustainable functioning of the respective market where sufficient competition is in place in order to balance the interests of involved stakeholders (telecom operators, content providers, business users, consumers, etc.). Another role of the provider may be to mitigate the effects of digital divide by making technology and the access to information accessible to anyone. For this reason, mitgating the market power accumluation of individual stakeholders is of central importance in order to lower the probability of unbeneficial or even defective strategies, e.g., collusion. This role is addressed in this section by reconsidering the separation of powers within the ecosystem.


Most operator services (such as SMS, telephony, etc.) are today provisioned by exactly the same entity as the infrastructure (providing the data connectivity). However, in a world of technological convergence this creates potential tensions between the interests of competing service providers and the operator. While content providers may have to pay for a preferential treatment of their traffic, if allowed by Net Neutrality regulations in the specific market, those classical services may have to be prioritized without additional charges in order maintain the known quality standards. Even if the providers had to split their revenues between service and infrastructure departments, the process would be intransparent. Referring to technological convergence, newly added capacity could be associated to specific services only. Of course, following a similar line of thought, this also opens the doors to prioritization business models. This technological convergences also stretches to the integration of unlicensed spectrum bands in the context of cellular services (e.g., via LTE-U). Subtly defining the licensed and unlicensed band cellular services to be separate products, discriminatory practices could be extended to determining whether a service is hosted only in unlicensed or licensed, or in either sort of spectrum bands.


Hence, operators may be able to prize their interests above others. For this reason, the essential prerequisite for a self-regulating market is the reasonable separation of powers, which starts by decoupling any kind of Internet-provided service from the actual infrastructure (as moderator of the publicly provided spectrum), i.e., totally independent corporations are required. Thus, operator-native voice services are treated equally to external services and both have to face the same charges (whether for basic termination and transmission or for priority handling). This yields the following benefits: 
\begin{inparaenum}
	\item Avoidance of unjustified preference of own services (not in accordance to market prices);
	\item Transparent overview of market prices and costs;
	\item Higher competition is established in each part of the value chain, i.e.,  in operator-services, infrastructures, contents, access business, etc.
\end{inparaenum}

Similar solutions have been introduced during the market liberalization phase in the railway (analogies exist in the area of price discrimination \cite{Odlyzko:2012tz}) and energy domains where infrastructure operations, e.g., rail investment and maintenance or Distribution System Operators (DSOs) respectively, have been decoupled from the end customer business. Technologically, this is enabled by recent efforts to virtualize essential network functions, as described in \cite{forde2011exclusive}.


	The same line of thought can also be applied to the utilization of spectrum, irrespectively of whether it is exclusively licensed or publicly shared. Whenever the utilization of unlicensed spectrum is left unregulated and is open for business practices of big players, constellations may arise that lead to a tragedy of commons \cite{hardin1968tragedy} as illustrated in \cite{zwickl2013wi}. Contrary, Ostrom et al. \cite{ostrom1999revisiting} argue that the direct access to common resources can be collectively managed whenever physical access to the resource and sound community relationships exist. While those conditions may be met for privately operated Wi-Fi hotspots, they can hardly be satisfied for larger networks with intransparent spectrum usage and inaccessible deployment locations for users.  For this reason, a regulated or self-regulating market has to protect the interest of the society and each individual. This is today targeted by common regulatory policies that only carefully grant exclusive spectrum access to ISPs under certain market penetration and quality targets. In other words, public spectrum has to be put into use for the society who owns it.	Another future regulation may have to focus on similar measures in the unlicensed spectrum in order to keep the competitive situation stable and the unlicensed spectrum open for non-commercial usages.

\begin{figure}[htbp]
\centering
\includegraphics[width=0.49\textwidth]{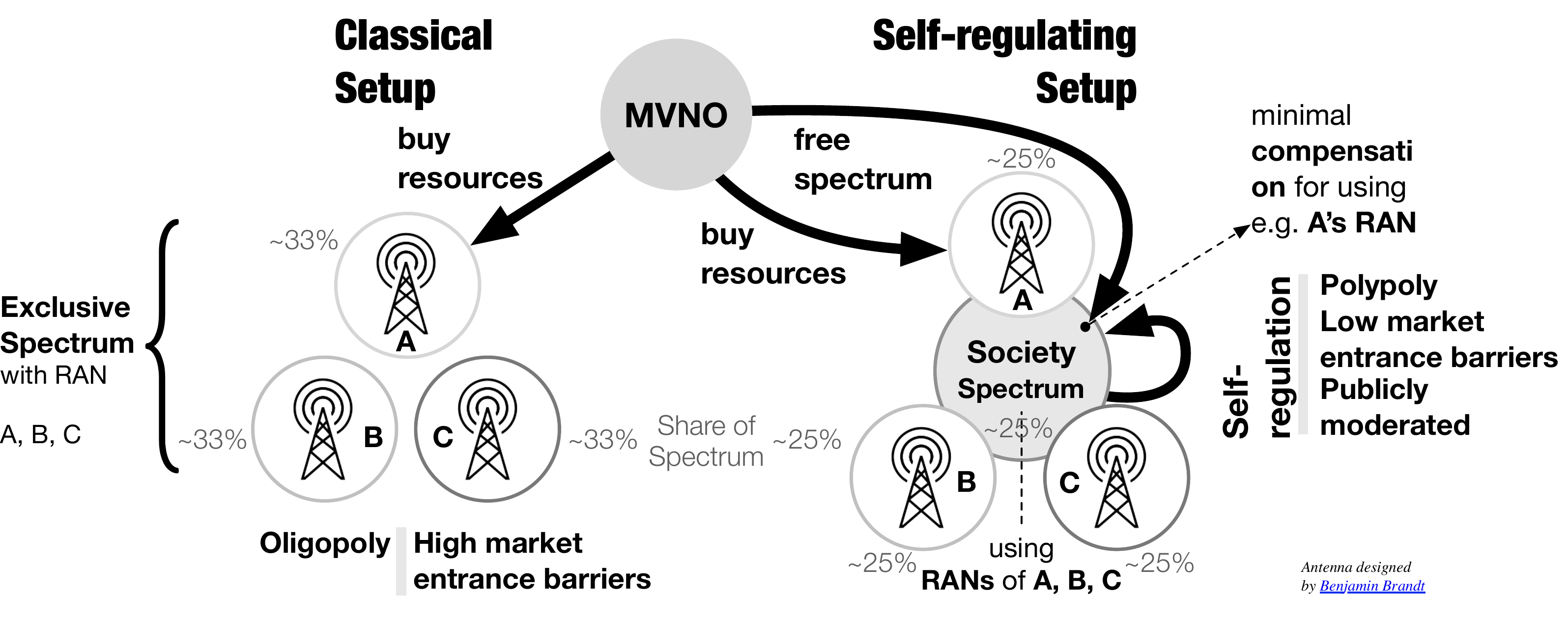}
\caption{From rigoros oligopoly regulations to self-regulated polypoly \& oligpoly spectrum markets.}
\label{fig:oldvsnewpicture}
\end{figure}

\section{Market Self-Regulation via Society Spectrum} \label{sec:market_design}


We will now construct a \textit{competitive self-regulating} \cite{ogus1995rethinking} cellular telecommunications market, following the stated separation of powers, by introducing the \textit{society spectrum}. The self-regulation model may best be classified as  ``Independent Agency-assisted Competition'' (where the regulator is the agency that represents the society for setting quality standards and legal conditions) where only limited constraints are posed to individual markets.

Today's spectrum is exclusively assigned to commercial network operators by using auction mechanisms (cf. Fig. \ref{fig:oldvsnewpicture}, \textit{ Classical Setup} and \cite{melody1980radio}). By meaningfully distributing spectrum among a number of network operators, a balanced market is targeted, which is attractive both for the society (other businesses, users, etc.) and the bidding operators. However, market consolidations, if approved, may substantially shift market powers, which may trigger actions by the regulators. The market entrance of MVNOs is often seen as measure for softly and dynamically adding competition towards obtaining a more fine-granular control over the market. Nevertheless, today's market setup requires rigorous regulatory actions both at the strategic (long run) and the operational (short run; day-to-day business) level.

In this light, the present work proposes an alternative spectrum management regime (cf. Fig.~\ref{fig:oldvsnewpicture}, \textit{Self-regulating Setup}), which envisions to focus the regulatory involvement on strategic measures, while operational control is replaced by a self-regulated regime. In particular our approach aims at relieving regulators from actively steering and moderating market entrances, market exits, and the redistribution of spectrum. Of course, rather than replicating the unlicensed spectrum organization (e.g., Wi-Fi) in the licensed spectrum, we intend to assure a long term functioning of the exclusive spectrum in coexistence to unlicensed spectrum solutions and as a \textit{happy medium} between today's licensed and unlicensed spectrum approaches.



\begin{figure}[htbp]
\centering
\includegraphics[width=0.50\textwidth]{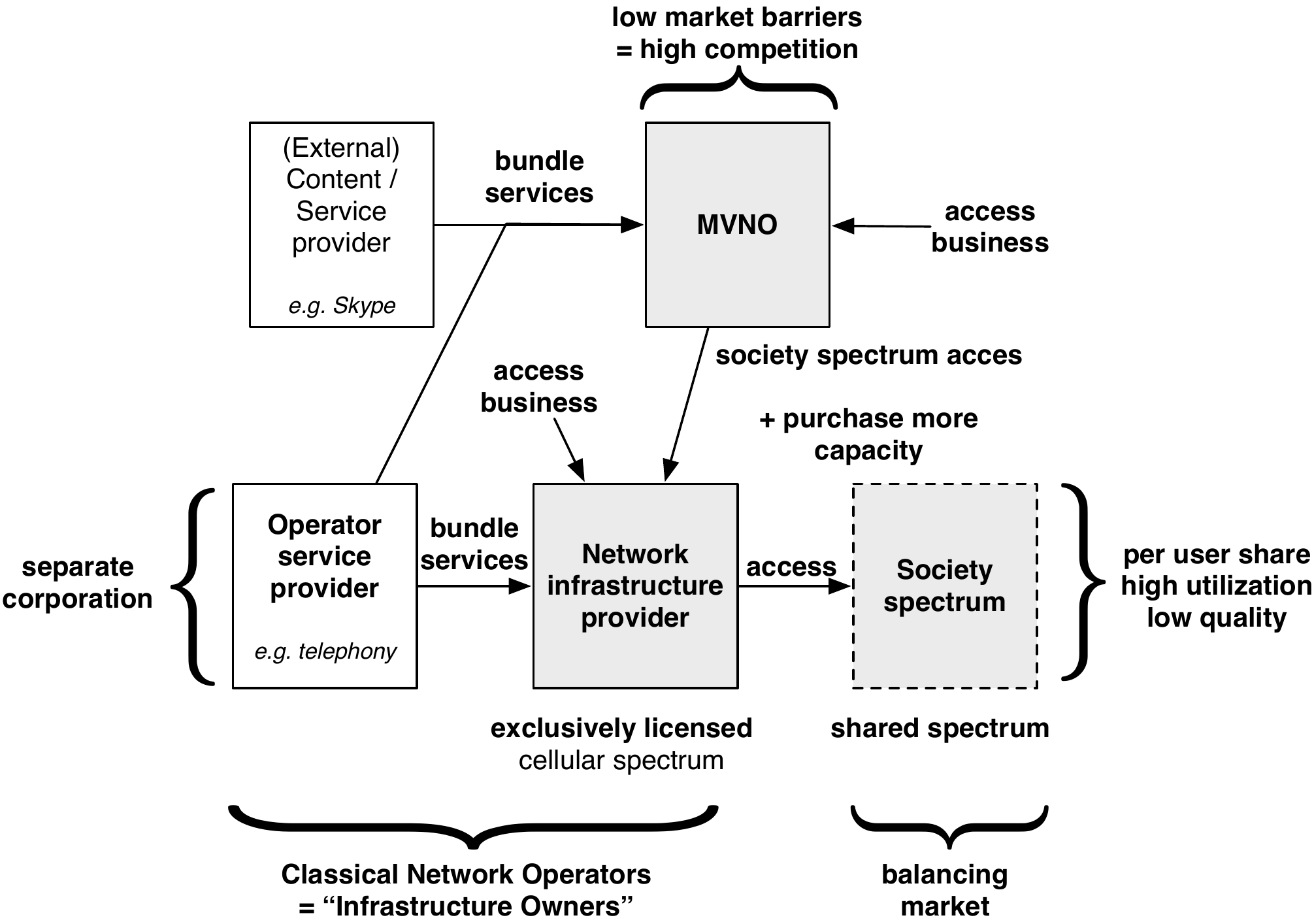}
\caption{A self-regulating cellular network market design.}
\label{fig:big_picture}
\end{figure}

	Towards obtaining this goal the society spectrum is defined as share $w$ of the overall available spectrum, which is available for auction. The society spectrum is shared among all MVNOs   and provided free of charge. For separation of powers reasons, classical operators can only access the society spectrum themselves by creating a separately organised subsidiary, i.e., an MVNO. 
	
	The factor $w$ represents a weighting factor for the functioning of the self-regulation. If $w$ is too low, consumer rights are endangered (due to insufficient competitive market pressure), but if $w$ is too high the entire industry is endangered (due to unprofitability). The paramterization of $w$ is subject to the considered market. For smaller countries such as Austria, the number of $4$ classical operators have added extensive market pressure, which may harm infrastructure investments in the long run. The reduction to $3$ operators, however, has led to a disproportionate price increase\footnote{\url{http://www.bwb.gv.at/Aktuell/Seiten/Telekombranchenuntersuchung-Zwischenstatus.aspx}. Last accessed: 2015-03-27}. Those extrema constitute the action space for the society spectrum, thus $\approx 25\%$ (of comparable spectrum bands or band bundles) could initially be assigned to the society spectrum in this case. The rest is split among 3 winning operators A, B and C. Based on functioning of the classical operator's market (A, B, C), the market pressure can be increased or lowered to almost one of those extrema. The factor $w$, hence, models the relief spectrum, where the level of competition moderates the usage of the relief valve between both kinds of spectrums. In other words, when providers $A$, $B$ and $C$ collude, the prices will go up or the quality will stagnate, which will lower the utility $\mathcal{U}^e$ for purchasing services from those providers. In today's regime we can assume that users will purchase a service as long as their utility $\mathcal{U}^e$ is positive. With the introduction of the society spectrum however, the utility gain $\mathcal{U}^e - \mathcal{U}^s$ needs to be positive. Thus, in addition to remaining price-quality considerations of consumers, the added value over the society spectrum utility $\mathcal{U}^e$ creates additional pressure. The fewer people recognise an added value of additional contracts, the lower will be the market success of $A$, $B$, $C$. The opposite development kicks-in when the quality in the society spectrum is low, while exclusive spectrum offers are attractive.
	
	
In an assessment on ``mission critical'' services \cite{forge2013}, the creation of dedicated infrastructure is associated to high investment costs and a common need for backhaul services. In analogy,  the revised regulatory regime presented in this work should limit the requirement for investments in separate infrastructure in order to keep market barriers low. Hence, the access to the society spectrum is enabled via the infrastructure of the winning operators ($A$, $B$, $C$ in our example) of the licensed spectrum auction, i.e., MVNOs are accessing the common spectrum via the RANs and backhaul networks of classical cellular operators---cf. Fig.~\ref{fig:big_picture}. Those cellular operators have to open their infrastructure (primarily their RANs and backhaul infrastructures) for MVNOs in order to be allowed to exclusively use their licensed spectrum. They may receive a fixed and low compensation payment, which corresponds to the line fees in the fixed line environment and may be strictly regulated. The access to the  \textit{society spectrum} is free of charge for MVNOs.

	
In our prior work on cellular traffic offload \cite{zwickl2013wi}, we have demonstrated that the Quality of Service (QoS) for commonly accessible resources will be strictly lower than for exclusively operated spectrum. In the example of Figure \ref{fig:oldvsnewpicture}, the quality of the publicly-moderated society spectrum will be lower than in the exclusive spectrum of A, B, C. Nevertheless, MVNOs may also distinguish themselves from other MVNOs by purchasing a slice of the licensed spectrum, as is common practice in today's telecommunications industry, in order to improve their QoS. Operator services may be bought from any external service provider that is capable of meeting the desired quality standards. This again entails a few notable advancements over the current state of non-self-regulatory control:
	\begin{inparaenum}
		\item The society spectrum provides natural control over the market, which shifts bigger market shares entirely to the society spectrum whenever the prices become unreasonable. Whenever the competition among MVNOs (solely in the unlicensed spectrum) becomes too challenging, MVNOs may exit the market (due to insufficient market prospects) or may start a quality strategy (e.g., by purchasing more licensed spectrum shares from network operators).
		\item Classical cellular operators (owning infrastructure) will have limited incentives to enter the low cost and low profit MVNO-business, hence strategies for intentional overload, such as envisioned in \cite{zwickl2013wi} for Wi-Fi unlicensed spectrum, can be mitigated or eliminated.
		\item Exclusively licensed spectrum will hold the premium traffic, while the low cost and seamless access to the society spectrum will substantially lower the Quality of Service (QoS) (cf. \cite{zwickl2013wi}). Hence, both societal needs and business consideration of RAN operators will be met.
		\item Very low market entrance barriers will probably provide the best approximation of a perfect competition that can be constructed in a telecommunications market.
		\item The new approach facilitates the modularity both on the technological and on the competitive axis, i.e., each separate service faces a separate competitive situation of directly interchangeable services and under transparent market conditions.
	\end{inparaenum}
	
\begin{figure}[htbp]
\centering
\includegraphics[width=0.40\textwidth]{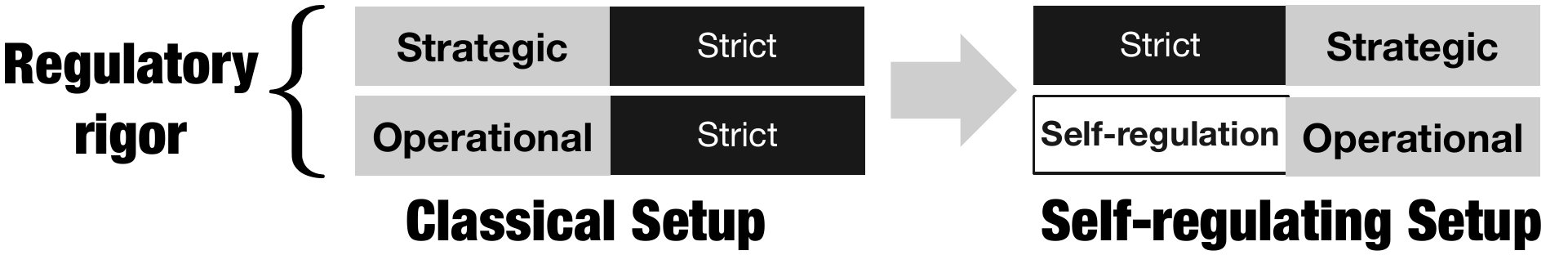}
\caption{Introducing self-regulation to the regulatory landscape.}
\label{fig:regulation}
\end{figure}

	
In terms of regulation, the strict strategic (long-term) and operational (short-term) regulation regime can be replaced by a less rigorous counterpart in our new concept---see Figure \ref{fig:regulation}. Due to the self-regulating properties, regulatory measures on the operational level can be reduced to a minimum. The incentives for A, B and C to misuse their market power, due to their exclusive access to some spectrum, is eliminated by the introduction of the society spectrum. Whenever product prices increase disproportionately (e.g., due to collusion), the society spectrum can relieve customers of some added pressure, i.e., some customers may terminate their contracts with exclusive cellular operators. Due to the increased load in the society spectrum and the falling revenues for operators A, B, and C, a counter-reaction will be triggered, i.e., an equilibrium will be found due to self-regulation. Of course, the proper strategic setup and the observation of illegal business practices will remain challenging tasks for the regulators.

Similarly, the unlicensed spectrum (mainly Wi-Fi) after the introduction of LTE-U may be moderated and regulated in a comparable fashion to the society spectrum. However, while the society spectrum is semi-orchestrated and semi-hierarchical (despite its self-regulatory nature), Wi-Fi bands should only be regulated whenever the competition for supremacy over resources becomes too fierce.

\section{Fairness Concepts} \label{sec:fairness}

A self-regulating system should further aim at assuring fair distribution of resources among users, which in the present case translates to a fair usage of spectrum. In the light of Net Neutrality discussions, each byte is supposed to have ``equal rights'' in the society spectrum, where no preferential treatment or other kind of discriminatory action against individuals, particular class of customers or services is allowed. Thus, all users are treated equally and assisted by a cooperative organization.

To this end, we recommend two complentary fair resource allocation principles, which could in practice, e.g., be realized based on a number of different resource sharing concepts and queuing techniques (see below):



\begin{itemize}
	\item \textbf{Mutual user fairness:} Based on the user base of the MVNO, the MVNO gets assigned a corresponding portion of the dedicated total of the society spectrum. In analogy, the more users are represented by a single MVNO, the higher the share of the dedicated society spectrum becomes for the physical network operator (e.g., A, B or C) hosting the MVNO (cf. \secref{sec:spectrum_access} for detailed as well as alternative realizations of this concept). The meaningful and fair distribution of resources among society spectrum users needs to be moderated by the MVNO and the by relevant regulatory frameworks, based e.g. on well-known queuing concepts like Weighted Fair Queueing (WFQ) \cite{wfq-acm, wfq-ieee} or Deficit Round Robin (DRR) \cite{drr-ton}. The work conservation property of WFQ and DRR thereby assures that no resources are left unutilized (i.e., idle) in the presence of users' traffic demands.
	\item \textbf{Spectrum conservation:}
	Whenever a provider experiences periods of underutilization (e.g., due to a lack of customer interest induced by inappropriate pricing) its licensed spectrum is dynamically added to the already existent society spectrum-pool. The proposed approach is work-conservative in that it will assure that idle licensed spectrum is always at disposal for society spectrum users. On a technical level, this scheme could be realized based on a simple priority concept in which the licensed spectrum users always assume absolute priority in the utilization of the licensed spectrum.  
	
\end{itemize}

The fair assignment of resources, in analogy to the \textit{mutual user fairness} principle, has extensively been studied in radio access technology literature. For example Xu \textit{et al.} \cite{xu2002dynamic} assess fair scheduling mechanisms for the dynamic bandwidth assignment and Quality of Service (QoS) differentiation in WCDMA-based 3G and 4G networks. While they specifically focus on the Generalized Processor Sharing (GPS) concept \cite{parekh1993generalized}, they also provide explicit reference to WFQ.  Multi-user fairness within Long Term Evolution (LTE) networks has been addressed in \cite{kwan2009proportional} by  specifying a proportional fairness scheduler. For unlicensed spectrum technologies, an architecture for a distributed flavour of WFQ has  been proposed in \cite{banchs2002distributed} for Wi-Fi networks following the IEEE 802.11 standards\footnote{The Working Group for WLAN Standards: \url{http://grouper.ieee.org/groups/802/11/}, last accessed: \today}.  







\section{Spectrum Assignment Models} \label{sec:spectrum_access}


Following the defined fairness concepts, the corresponding assignment of society spectrum shares needs to be put into the context of the operator landscape. This section will discuss models that translate the user's fair share of the society spectrum to solutions involving rationally-acting operators. A specific challenge lies in the separation of the society spectrum from each provider's own spectrumin order to avoid cheating practices, while using the same RAN. The following flavors for realizing the access to the society spectrum exist:

\begin{enumerate}
\item \textbf{Chaotic organization:} Like the unlicensed spectrum (e.g., Wi-Fi) the co-usage of the society spectrum may not be regulated or organized in the classical sense. In our model, every network operator would provide access to the society spectrum but would not aim at controlling the quality therein. Hence, apart from the infrastructure co-use, no added value will be provided to end users over today's unlicensed spectrum solutions.


\item \textbf{Billing:} The society spectrum may be \textit{virtual} and added to the capacity of each operator's exclusively owned part. For every user the MVNO receives a share of the society spectrum. Smart billing approaches may be used to determine whether the usage is within the defined limits for the designated society spectrum share. For excess traffic, normal MVNO fees have to be paid in order to compensate for the additional spectrum usage. However, the associated billing, monitoring and verification process would be highly complex. Hence, a practical usage appears to be unrealistic.

\item \textbf{New \textit{virtual} operator:} A virtual operator can be created in order to clearly keep apart the different kinds of resources. The virtual operator is probably jointly owned by all other operators (whether MVNOs or providers with own RAN) and managed by a public non-profit organization. In this paradigm, the costs are equally shared among all market participants and all resources are provided with low market barriers, whereby the infrastructures of the RAN operators are reused to the greatest extent possible. Due to the strict separation of market players and spectrum resources, the appropriate usage of the society spectrum is easy to monitor. However, from the current technological point of view, session continuity could become problematic whenever additional resource are required for a particular session on a micro-flow level, i.e., when users dynamically need to utilize exclusive spectrum in order to satisfy the bandwidth demands which exceed their individual society spectrum shares.




\item \textbf{Per operator society spectrum:} Each physical RAN operator runs a specific share of the society spectrum. The size of its spectrum share corresponds to the sum of end users represented by the  MVNOs hosted by the physical operator and the number of the physical operator's customers themselves. In other words, the size of the society spectrum share for each physical RAN operator directly corresponds to the number of users that utilize the RAN operator's infrastructure. This scheme is able to provide fair spectrum shares to the individual  users, provided that the size of the society spectrum share for each physical operator can regularly (e.g., once per month) be adjusted to the number of users it effectively hosts. Compared to the \textit{new virtual operator} flavor, this model can seamlessly assure dynamic spectrum assignment to the individual microflows whenever the users' society spectrum shares cannot accommodate their traffic demands.



\end{enumerate}


The fair assignment of spectrum may face additional difficulties like the following: Users may create several contracts with different or the same provider in order to obtain a larger share of the society spectrum. As this behaviour represents a \textit{dominant strategy}, the assignment of a fair share of society spectrum to human users should be associated with an identity check. To this end, the regulator could for example provide each human user with a personal and unique access code to the society spectrum, which the user can simultaneously use only via a single physical network infrastructure.

\section{Concluding Remarks}
\label{conclusions}

The present work has proposed a novel ``self-regulating'' spectrum market design in which essential coordination functions known from the current cellular paradigms are kept unchanged, whereas the regulator is simultaneously relieved of excessive day-to-day operations' oversight. Self-regulation in our work refers to the automatic balancing of interests due to an intelligent distribution of market powers. This has been obtained by introducing the concept of a ``society spectrum'', which guarantees a fair fulfillment of minimal connectivity demands for all users at very low costs. We have further illustrated the practicability of our design by constructing a relationship to fair radio access principles, e.g., based on WFQ scheduling, and by illustrating the assignment of spectrum shares to individual operators in correspondence to the represented user base. In future works, we target the detailed linkage to cognitive radio solutions and the conduction of corresponding simulations to study the market regime stability.




\bibliographystyle{IEEEtran}    
\balance
\bibliography{main} 

\begin{thebibliography}{10}
\providecommand{\url}[1]{#1}
\csname url@samestyle\endcsname
\providecommand{\newblock}{\relax}
\providecommand{\bibinfo}[2]{#2}
\providecommand{\BIBentrySTDinterwordspacing}{\spaceskip=0pt\relax}
\providecommand{\BIBentryALTinterwordstretchfactor}{4}
\providecommand{\BIBentryALTinterwordspacing}{\spaceskip=\fontdimen2\font plus
\BIBentryALTinterwordstretchfactor\fontdimen3\font minus
  \fontdimen4\font\relax}
\providecommand{\BIBforeignlanguage}[2]{{%
\expandafter\ifx\csname l@#1\endcsname\relax
\typeout{** WARNING: IEEEtran.bst: No hyphenation pattern has been}%
\typeout{** loaded for the language `#1'. Using the pattern for}%
\typeout{** the default language instead.}%
\else
\language=\csname l@#1\endcsname
\fi
#2}}
\providecommand{\BIBdecl}{\relax}
\BIBdecl

\bibitem{croicioni}
P.~Croicioni, ``{Net Neutrality: Few thoughts and many questions},'' {Ofcom
  Whitepaper}, 2008.

\bibitem{borenstein1990airline}
S.~Borenstein, ``Airline mergers, airport dominance, and market power,''
  \emph{The American Economic Review}, pp. 400--404, 1990.

\bibitem{zwickl2013wi}
P.~Zwickl, P.~Fuxjaeger, I.~Gojmerac, and P.~Reichl, ``Wi-fi offload: Tragedy
  of the commons or land of milk and honey?'' in \emph{Personal, Indoor and
  Mobile Radio Communications (PIMRC Workshops), 2013 IEEE 24th International
  Symposium on}.\hskip 1em plus 0.5em minus 0.4em\relax IEEE, 2013, pp.
  148--152.

\bibitem{ogus1995rethinking}
A.~Ogus, ``Rethinking self-regulation,'' \emph{Oxford Journal of Legal
  Studies}, pp. 97--108, 1995.

\bibitem{kpmg96}
{KPMG}, ``{Public Policy Issus Arising from Telecommunications and Audiovisual
  Convergence},'' {Report for the European Commission}, 1996.

\bibitem{blackman1998convergence}
C.~R. Blackman, ``Convergence between telecommunications and other media: How
  should regulation adapt?'' \emph{Telecommunications policy}, vol.~22, no.~3,
  pp. 163--170, 1998.

\bibitem{swire1997markets}
P.~Swire, ``Markets, self-regulation, and government enforcement in the
  protection of personal information, in privacy and self-regulation in the
  information age by the us department of commerce.'' \emph{Privacy and
  Self-Regulation in the Information Age by the US Department of Commerce},
  1997.

\bibitem{sinclair1997self}
D.~Sinclair, ``Self-regulation versus command and control? beyond false
  dichotomies,'' \emph{Law \& Policy}, vol.~19, no.~4, pp. 529--559, 1997.

\bibitem{king2000industry}
A.~A. King and M.~J. Lenox, ``Industry self-regulation without sanctions: The
  chemical industry's responsible care program,'' \emph{Academy of management
  journal}, vol.~43, no.~4, pp. 698--716, 2000.

\bibitem{melody1980radio}
W.~H. Melody, ``Radio spectrum allocation: role of the market,'' \emph{The
  American Economic Review}, pp. 393--397, 1980.

\bibitem{faulhaber2003spectrum}
G.~R. Faulhaber and D.~J. Farber, ``Spectrum management: Property rights,
  markets, and the commons,'' \emph{Rethinking rights and regulations:
  institutional responses to new communication technologies}, pp. 193--226,
  2003.

\bibitem{niyato2008market}
D.~Niyato and E.~Hossain, ``Market-equilibrium, competitive, and cooperative
  pricing for spectrum sharing in cognitive radio networks: Analysis and
  comparison,'' \emph{Wireless Communications, IEEE Transactions on}, vol.~7,
  no.~11, pp. 4273--4283, 2008.

\bibitem{chapin2007}
J.~M. Chapin and W.~H. Lehr, ``The path to market success for dynamic spectrum
  access technology,'' \emph{IEEE Communications Magazine}, p.~97, 2007.

\bibitem{rodriguez2005market}
V.~Rodriguez, K.~Moessner, and R.~Tafazolli, ``Market-driven dynamic spectrum
  allocation: Optimal end-user pricing and admission control for cdma,''
  \emph{Proc. 14th European information society technologies (IST) mobile and
  wireless communications summit. Dresden}, 2005.

\bibitem{holland2012pluralistic}
O.~Holland, L.~De~Nardis, K.~Nolan, A.~Medeisis, P.~Anker, L.~F. Minervini,
  F.~Velez, M.~Matinmikko, and J.~Sydor, ``Pluralistic licensing,'' in
  \emph{Dynamic Spectrum Access Networks (DYSPAN), 2012 IEEE International
  Symposium on}.\hskip 1em plus 0.5em minus 0.4em\relax IEEE, 2012, pp. 33--41.

\bibitem{forde2011exclusive}
T.~K. Forde, I.~Macaluso, and L.~E. Doyle, ``Exclusive sharing \&
  virtualization of the cellular network,'' in \emph{New Frontiers in Dynamic
  Spectrum Access Networks (DySPAN), 2011 IEEE Symposium on}.\hskip 1em plus
  0.5em minus 0.4em\relax IEEE, 2011, pp. 337--348.

\bibitem{Odlyzko:2012tz}
A.~Odlyzko, ``{The Evolution of Price Discrimination in Transportation and its
  Implications for the Internet},'' \emph{Review of Network Economics}, vol.~3,
  no.~3, pp. 1--24, Mar. 2012.

\bibitem{hardin1968tragedy}
G.~Hardin, ``The tragedy of the commons,'' \emph{science}, vol. 162, no. 3859,
  pp. 1243--1248, 1968.

\bibitem{ostrom1999revisiting}
E.~Ostrom, J.~Burger, C.~B. Field, R.~B. Norgaard, and D.~Policansky,
  ``Revisiting the commons: local lessons, global challenges,'' \emph{science},
  vol. 284, no. 5412, pp. 278--282, 1999.

\bibitem{forge2013}
S.~Forge, R.~Horvitz, and C.~Blackman, ``{Is Commercial Cellular Suitable for
  Mission Critical Broadband?}'' {A study prepared for the European Commission
  DG Communications Networks, Content \& Technology by SCF Associated Ltd},
  2013.

\bibitem{wfq-acm}
\BIBentryALTinterwordspacing
A.~Demers, S.~Keshav, and S.~Shenker, ``Analysis and simulation of a fair
  queueing algorithm,'' \emph{SIGCOMM Comput. Commun. Rev.}, vol.~19, no.~4,
  pp. 1--12, Aug. 1989. [Online]. Available:
  \url{http://doi.acm.org/10.1145/75247.75248}
\BIBentrySTDinterwordspacing

\bibitem{wfq-ieee}
A.~Parekh and R.~Gallager, ``A generalized processor sharing approach to flow
  control in integrated services networks: the single-node case,''
  \emph{Networking, IEEE/ACM Transactions on}, vol.~1, no.~3, pp. 344--357, Jun
  1993.

\bibitem{drr-ton}
M.~Shreedhar and G.~Varghese, ``{Efficient Fair Queuing Using Deficit
  Round-Robin},'' \emph{IEEE/ACM Transactions on Networking}, vol.~4, no.~3,
  pp. 375--385, June 1996.

\bibitem{xu2002dynamic}
L.~Xu, X.~Shen, and J.~W. Mark, ``Dynamic bandwidth allocation with fair
  scheduling for wcdma systems,'' \emph{Wireless Communications, IEEE}, vol.~9,
  no.~2, pp. 26--32, 2002.

\bibitem{parekh1993generalized}
A.~K. Parekh and R.~G. Gallager, ``A generalized processor sharing approach to
  flow control in integrated services networks: the single-node case,''
  \emph{IEEE/ACM Transactions on Networking (ToN)}, vol.~1, no.~3, pp.
  344--357, 1993.

\bibitem{kwan2009proportional}
R.~Kwan, C.~Leung, and J.~Zhang, ``Proportional fair multiuser scheduling in
  lte,'' \emph{Signal Processing Letters, IEEE}, vol.~16, no.~6, pp. 461--464,
  2009.

\bibitem{banchs2002distributed}
A.~Banchs and X.~Perez, ``Distributed weighted fair queuing in 802.11 wireless
  lan,'' in \emph{Communications, 2002. ICC 2002. IEEE International Conference
  on}, vol.~5.\hskip 1em plus 0.5em minus 0.4em\relax IEEE, 2002, pp.
  3121--3127.

\end{thebibliography}

\end{document}